\documentclass[a4paper]{article}
\usepackage[english]{babel}
\usepackage[utf8]{inputenc}

\usepackage{fix-cm}
\usepackage{amssymb}
\usepackage{amsmath}
\usepackage{graphicx}
\usepackage{subfigure}
\usepackage{makeidx}
\usepackage{multicol}
\usepackage{bbm}
\usepackage[numbers]{natbib}
\usepackage{nameref}
\usepackage{url}
\usepackage{hyperref}

\newcommand{\Reals}{\mathbb{R}}

\newcommand{\Prob}{\mathbb{P}}
\newcommand{\E}{\mathbb{E}}

\newcommand{\indep}{\perp \!\!\! \perp}

\DeclareMathOperator*{\pa}{pa}
\DeclareMathOperator*{\An}{An}
\DeclareMathOperator*{\an}{an}
\DeclareMathOperator*{\RV}{RV}

\usepackage{amsthm}
\newtheorem{theorem}{Theorem}[section]
\newtheorem{definition}{Definition}[section]
\newtheorem{example}{Example}[section]

\makeindex
\usepackage{tikz-cd}
\newcounter{mylabels}
\tikzset{label label/.code={\stepcounter{mylabels}},
label labels/.style={every label/.append style={label label,
    alias=mylabel-\number\value{mylabels}}},
fit in labels/.style={label labels,
execute at begin picture={\edef\myfirstlabel{\the\numexpr\value{mylabels}+1}},
execute at end picture={\edef\mylastlabel{\the\numexpr\value{mylabels}}
\path foreach \X in {\myfirstlabel,\the\numexpr\myfirstlabel+1,...,\mylastlabel}
{foreach \Anchor in {north,south,west,east} 
 {(mylabel-\X.\Anchor)}};}}}
\tikzcdset{mdiagram/.style={every arrow/.append style={-latex,semithick},
    /tikz/column sep=2em,/tikz/row sep=2em,
    /tikz/.cd,dot/.style={circle,fill,inner sep=2pt},
    every label/.append style={overlay},fit in labels},
    }

\title{Causality and extremes}

\author{Valérie Chavez--Demoulin$^{a}$, Linda Mhalla$^{b}$ \\
        \small $^{a}$HEC, University of Lausanne, Switzerland, \href{mailto:valerie.chavez@unil.ch}{valerie.chavez@unil.ch} \\
        \small $^{b}$EPFL, Switzerland, \href{mailto:linda.mhalla@epfl.ch}{linda.mhalla@epfl.ch} \\
}
\date{\today}

\begin{document}
\maketitle
\begin{abstract} 
\noindent In this work, we summarize the state-of-the-art methods in causal inference for extremes. In a non-exhaustive way, we start by describing an extremal approach to quantile treatment effect where the treatment has an impact on the tail of the outcome. Then, we delve into two primary causal structures for extremes, offering in-depth insights into their identifiability.  Additionally, we discuss causal structure learning in relation to these two models as well as in a model-agnostic framework. To illustrate the practicality of the approaches, we apply and compare these different methods using a Seine network dataset. This work concludes with a summary and outlines potential directions for future research.\end{abstract}

\section{Introduction}\label{ch19:intro}
Until now, the fields of causal inference \cite{Spirtes, pearl_09,Peters2017} and extreme value theory \cite{emb1997,Beirlant_2004,deHaan.Ferreira:2006,Resnick:2006} have mostly been separate. In order to understand the causal mechanisms and the consequences of extremes, these fields must be brought together. There are different schools of causal inference with the two prominent ones being the potential outcome models \cite{Rubin_1974} and causal graphs (along with the do-operator) \cite{pearl_09}. Despite philosophical differences between the schools, the fundamental task of causal inference can be described as comparing outcomes under different regimes (treatments). Here, comparison of outcomes can take different forms. For instance, one can distinguish between two lines of work in causality: estimation and identifiability of causal effects and causal structure learning. Works on identifiability and estimation of causal effects include the instrumental variables approaches \cite{imbens1994} and covariate adjustments through the backdoor criterion \cite{Pearl1993} and adjustment criterion \cite{perkovic2018}. Methods inferring the causal structure among variables, often represented by a directed acyclic graph (DAG), are known as observational causal discovery methods \cite{lingam,hoyer2009,spirtes2016}. Depending on the set of the causal assumptions (sufficiency, faithfulness, etc) that one assumes, there exists a variety of causal inference methods including the constraint-based methods (based on conditional independencies between variables) and the score-based methods (based on a greedy search over the space of possible completed partially DAGs) \cite{Chickering}. Alternatively, \cite{mooij2010} proposed to use the postulate of the independence of cause and generating mechanism to formulate the problem of causal discovery through asymmetries in the Kolmogorov complexities of the marginal and the conditional distributions.

\subsection{Overview}
While the field of causality is well-established, it has mostly focused on causal effects on the mean of the outcome variable(s). This is a fundamental issue in many fields of science such as climatology where extreme event attribution deals with causal links between climate forcings and observed responses with the aim of attributing likely causes for a detected climate change \cite{hannart2016, kiriliouk_Naveau}. Bridging causal inference with extreme value theory (EVT) has thus been recently considered in an attempt to benefit from the solid foundation of EVT that deals with scarcity of observations and allows for extrapolation. Methods for causality of extremes from observational data have been proposed and we will cover some of them in this work.

\subsection{Illustrative dataset: description and motivation}
\label{ch19:sec:Seine}
\begin{figure}[htp]
    \centering
\includegraphics[width=8cm]{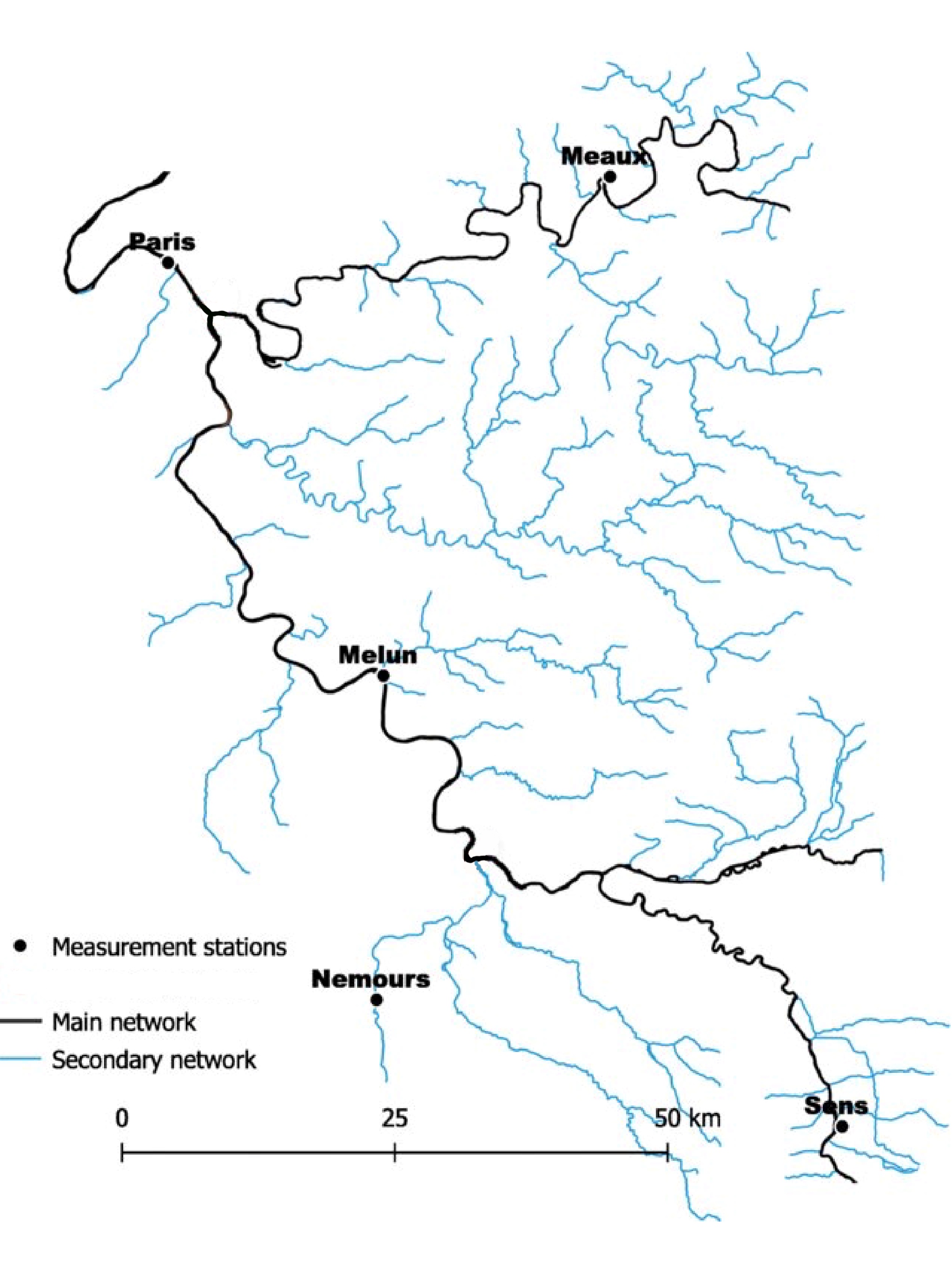}
\caption{Topographic map of part of the Seine network showing five sites along the Seine (Melun, Paris) or on its tributaries (Sens, Nemours, Meaux). Adapted from \cite{Asenova2021}.}
\label{ch19:fig:Seine}
\end{figure}

The Seine, a 774.76 km long river, rises at 471 meters above sea level on the Mont Tasselot in the C\^ote d’Or region of Burgundy.  It has a general orientation from south-east to north-west. Figure \ref{ch19:fig:Seine} shows part of the Seine network studied in \cite{Asenova2021}. The river goes through {\bf Melun} as its trenched valley crosses the \^Ile-de-France toward {\bf Paris}. {\bf Sens} is on the left-bank tributary of the Seine, called Yonne, a 292 km long river. {\bf Nemours} is on the left tributary of the Seine, called Loing, a 143 km long river and {\bf Meaux} is on the eastern tributary of the Seine, called Marne, a 514 km long river in the area east and southeast of Paris. 
Water level data is highly valuable in the literature of extreme event causality especially when the methodology does not depend on the time lag between two stations, i.e., on the direction of time. The analysis of water flows in the Bavarian Danube can be found in \cite{Mhalla_2020} and \cite{Gnecco_2021}, while analysis of water flows in Swiss rivers is discussed in \cite{Pasche_2022}. We assume that the ground truth is dictated by the physical orientation of the network. For instance, extreme water levels at station A would cause extreme water levels at station B if A is located upstream of B. The strength of this causal link, and thus the evidence of causation, may depend on various characteristics, such as whether A is situated on a tributary of the river where B flows, the distance between the two stations, and the size of the catchments. In Section \ref{ch19:sct:Seine}, we illustrate different methods of causal discovery for extreme water level on the Seine river. 

The rest of the paper is structured as follows. In Section \ref{ch19:quantile_effect} we describe an extremal approach of quantile treatment effect where the treatment has an impact on the tail of the outcome. The two main causal structures for extremes are described in Section \ref{ch19:causal_structures} with details about their identifiability in Section \ref{ch19:sct:identifiabiliy}. Consideration of the  causal structure learning related to the two models are exposed in Section \ref{ch19:sct:Learning}. Finally, the application and comparison of these methods on the Seine dataset are described in Section \ref{ch19:sct:Seine}. We conclude and mention potential avenues for future research in Section \ref{ch19:sct:Conclusion}.
%\section{Extremal causal effect estimation} 
\section{Quantile treatment effect: an extremal approach} 
\label{ch19:quantile_effect}
Let $Y$ denote the outcome of interest and $D$ a binary treatment or policy, or more generally the intervention. The Neyman--Rubin \cite{Neyman_1923,Rubin_1974} counterfactual framework provides a definition of causality based on comparing two competing potential outcomes: the outcome that occurred and the outcome that would have occurred under a different treatment. This definition is formalised under the concept of the \textit{average treatment effect}
\begin{equation*}
    \text{ATE} = \E \{ Y(1) \} - \E\{Y(0)\},
\end{equation*}
where $Y(1)$ is the potential outcome under treatment and $Y(0)$ the potential outcome without treatment. The popularity of this approach stems mainly from the few assumptions it requires to ensure identifiability of the ATE. For instance, the structure of causal mechanism linking the treatment to the outcome does not need to be specified and there are no distributional assumptions on the outcome of interest. 

Within this classical framework of causality, \cite{deuber2023} propose to study the size of a causal effect related to extremes. While a treatment might have a causal impact on the average values of a target outcome, the interest here lies in assessing the causal impact of such treatment on the high quantiles of the outcome. In the context of climate change, one might want to assess the causal impact of anthropogenic forcing on return levels of extreme precipitations, that might exceed the range of historical values. Thus, \cite{deuber2023} combine extreme value theory for extrapolation and counterfactual causal framework for measuring causal effects.

The causal impact on extremes is measured by the $\tau-$\textit{quantile treatment effect}
\begin{equation*}
    \text{QTE}(\tau) := q_1(\tau) - q_0(\tau),
\end{equation*}
where $q_i(\tau)$ is the $\tau$-quantile of the potential outcome $Y(i)$, for $\tau \in (0,1)$ and more precisely $\tau \approx 1$ or $\tau \approx 0$. To ensure identifiability of the QTE, the commonly-made assumption of unconfoundedness, i.e., $(Y(0),Y(1)) \indep D \vert X$, for a set of observed confounders $X \in \mathbb{R}^r$, is assumed. \cite{Firpo_2007} introduced estimators of the quantiles $q_i(\tau)$, for fixed $\tau \in (0,1)$ and hence of the QTE, where adjustments for confounding are made based on the propensity score $\Pi(x) := \Pr(D=1 \vert X=x)$ \cite{Rosenbaum_1983}. Asymptotic normality of the resulting estimate of the QTE are derived by \cite{Firpo_2007} when $\tau$ is fixed, and by \cite{Zhang_2018} under changing levels $\tau_n$. In the latter, as the sample size tends to infinity, the sequence of intermediate levels $\tau_n$ is assumed to converge to zero and the expected number of exceedances of the $(1-\tau_n)$-quantile either tends to infinity ($n \tau_n \rightarrow \infty$) or to a strictly positive quantity ($n \tau_n \rightarrow d >0$). In the context of extremes where extrapolation beyond observed values is necessary, one can reasonably assume that the distributions of the potential outcomes $Y(0)$ and $Y(1)$ are heavy-tailed and rely on the theory of regular variation to perform quantile extrapolation. More precisely, let $F_i$ denote the CDF of $Y(i)$, $i=0,1$, and suppose that they are heavy-tailed, i.e., 
\begin{equation*}
    1-F_i(y) \sim \ell_i(y) y^{-1/\xi_i}, \quad y \rightarrow \infty,
\end{equation*}
where $\xi_i$ is the extreme value index and $\ell_i$ a slowly varying function. Then, for an extreme quantile level $p_n \geq 0$, such that $p_n < \tau_n \rightarrow 0$ as $n \rightarrow \infty$, quantile extrapolation is performed as follows
\begin{equation*}
    q_i(1-p_n) \approx q_i(1-\tau_n) \bigg ( \dfrac{\tau_n}{p_n} \bigg) ^{\xi_i}.
\end{equation*}
\cite{deuber2023} rely on such extrapolation to modify the adjusted estimator of the QTE introduced by \cite{Firpo_2007} such that causal effects of a treatment on extreme quantiles, and no longer intermediate quantiles, can be measured. Their extremal QTE is thus approximated by
\begin{equation*}
    \text{QTE} (1-p_n) = q_1(1-p_n) - q_0(1-p_n) \approx q_1(1-\tau_n) \bigg ( \dfrac{\tau_n}{p_n} \bigg) ^{\xi_1} - q_0(1-\tau_n) \bigg ( \dfrac{\tau_n}{p_n} \bigg) ^{\xi_0}.
\end{equation*}
Therefore, estimation of the extremal QTE requires estimates of the extreme value indices of the potential outcomes' distributions. In the same spirit as the adjustment made by \cite{Firpo_2007} to account for observed confounders, an adjustment of the Hill estimator \cite{Hill_1975} based on inverse propensity score weighting is proposed. For a sample of $n$ independent observations $(Y_j, D_j, X_j)_{j=1}^n$ of the triplet $(Y,D,X)$, the resulting causal Hill estimators are defined as

\begin{align}
    \hat{\xi}_1^H &:=  \dfrac{1}{n \tau_n} \sum_{j=1}^n [ \log(Y_j) - \log\{ \hat{q}_1(1-\tau_n)\}] \dfrac{D_i}{\hat{\Pi}(X_j)} \mathbbm{1}_{Y_j > \hat{q}_1(1-\tau_n)},\notag 
    \\
    \hat{\xi}_0^H &:=\dfrac{1}{n \tau_n} \sum_{j=1}^n [ \log(Y_j) - \log\{ \hat{q}_0(1-\tau_n)\}] \dfrac{D_i}{1-\hat{\Pi}(X_j)} \mathbbm{1}_{Y_j > \hat{q}_0(1-\tau_n)},\notag
\end{align}
where $\hat{q}_i(1-\tau_n)$ are the adjusted (intermediate) quantile estimates of \cite{Firpo_2007} and $\hat{\Pi}$ is the estimate of the propensity score obtained by the sieve method; see \cite{Firpo_2007, Zhang_2018} for details. The extremal QTE estimator is then derived by plugging the causal Hill estimators $\hat{\xi}_i^H$ and the adjusted quantile estimates $\hat{q}_i(1-\tau_n)$. Under certain assumptions controlling the tail behaviours of the potential outcome distributions and their regularity, \cite{deuber2023} show asymptotic normality of the causal Hill estimators and the extremal QTE estimators. Finally, the authors derive a consistent estimator of the asymptotic variance such that statistical inference based on the asymptotic distribution, can be conducted.

The developed methodology targets settings where exposure to a treatment has an impact on the tail of an outcome of interest. For instance, the authors consider the causal effect of education on wages, a variable known for its heavy tails. While such settings are quite general, they are not applicable when extreme events are inherently multivariate. This is the case of the illustrative dataset of Section~\ref{ch19:sec:Seine} where a graph structure would be more adequate to capture interactions and causal effects between the sites.

%%%%%%%%%%%%%%%%%%%%%%%%%%%%%%%%%%%%%%%%%%%%%%%%%%%%%%%%%%%%
\section{Causal structures for extremes}
\label{ch19:causal_structures}
A general theory of causation called Structural Causal Model (SCM) \cite[Section 1.4]{pearl_09} is a combination of features from Structural Equation Models (SEM) \cite{Goldberger_73}, the framework of potential outcomes \cite{Rubin_1974}, and graphical models \cite{lauritzen_96}, offering probabilistic approaches to causation. Under the setting of recursive SEMs, a causal structure among several variables $\mathbf{X}=(X_1,\dots,X_d)^\top$ is represented by a directed acyclic graph (DAG), denoted $\mathcal{G}$, in which the set of nodes $V:=\{1,\dots,d\}$ represents the random variables and the set of directed edges $E$ the direct causal effects. We say that $i\in V$ is an ancestor of $j\in V$ in $\mathcal{G}$, if there exists a directed path from $i$ to $j$. We say that $j\in V$ is a descendant of $i\in V$ in $\mathcal{G}$, if there exists a directed path from $i$ to $j$. The set of descendants of $i$ is denoted by $De(i)$. The set of the ancestors of $j$ is denoted by $\An(j)$, and we define $\an(j):=\An(j)\setminus\{j\}$. If $\An(i)\cap\An(j)=\emptyset$, there is no causal link between $X_i$ and $X_j$. If there exist directed paths from $i$ to $j$ and from $i$ to $k$ in $\mathcal{G}$ that do not include $k$ and $j$, respectively, we say that $X_i$ is a confounder of $X_j$ and $X_k$.
%Two nodes $i,j$ connected by an edge in the skeleton of $\mathcal{G}$ are said to be adjacent. A triple of nodes $(i, j, k)$ is an unshielded triple if $i$ and $j$ are adjacent to $k$ but $i$ and $j$ are not adjacent. An unshielded triple $(i, j, k)$ is called a v-structure if $i$ is a direct ancestor (a parent) of $k$ and $j$ is a direct ancestor (a parent) of $k$. In this case, $k$ is called a collider. Two nodes $i,j$ are d-separated given $S\subset V$ if every path between $i$ and $j$ contains a noncollider that is in $S$ or a collider that is neither in $S$ nor an ancestor of a node in $S$.
The recursive SEM has the form
\begin{equation}
\label{ch19:eq:SEM}
X_j = f_j(X_{\pa(j)},\varepsilon_j), \quad j = 1,\ldots,d,
\end{equation}
where $\pa(j)\subseteq V$ is the set of parents (direct ancestors) of $j$, $f_j$ is a real-valued measurable function and $\varepsilon_1,\ldots, \varepsilon_d$ are independent noise variables. The distribution of $\mathbf{X}$ is uniquely defined by the distributions of the noise variables of the recursive SEM. 
%Denoting by $\rm{nd}(j)$ the non-descendants of node $j$, we have 
%$$X_j \indep X_{\rm{nd}(j)\backslash pa(j)} \mid X_{\pa(j)}, j = %1,\ldots,d,$$ 
%and we say that the distribution of $\mathbf{X}$ is Markov relative to $\mathcal{G}$.
%Under the Markov assumption, if all conditional (in-)dependencies can be read-off from the DAG $\mathcal{G}$ using the so-called d-separation rule, we say that the distribution of $\mathbf{X}$ is faithful to the DAG $\mathcal{G}$.  
The link between the DAG properties and the corresponding distributions is typically used for structure learning from observational data. 

%For any $i,j\in V$ we let $\beta_{i\rightarrow j}$ denote the sum of the products of the causal weights along the distinct directed paths from vertex $i$ to vertex $j$; we set $\beta_{j\rightarrow j}:=1$ and $\beta_{i\rightarrow j}:=0$ if $i\notin\An(j,G)$.

\subsection{Definitions and representations}
The majority of the literature related to causal structure model focuses on the bulk of the distribution. Often based on the Gaussian distribution, the related models usually lead to underestimation of extreme risks. Two recursive SEM \eqref{ch19:eq:SEM} are convenient for describing causal mechanisms apparent at the level of the tails of distributions: one supposes that the $f_j$ are linear and another one considers  the $f_j$ to be max-linear. We describe both below.
\begin{definition}\label{ch19:df:scm}
Consider a set of random variables $\mathbf{X}$. The following relation
\begin{equation}
\label{ch19:SCM}
    X_j:=\sum_{k\in \pa(j)}\beta_{jk}X_k+\varepsilon_j, \quad j\in V,%:=\{1,\dots,d\},
\end{equation}
  where 
  %$\pa(j)\subseteq V$ is the set of parents of $j$, 
  $\beta_{jk}\in\Reals\setminus\{0\}$ is the causal weight of node $k$ on node $j$, and $\varepsilon_1,\dots,\varepsilon_d$ are jointly independent noise variables, defines a linear structural causal model (LSCM) with associated directed acyclic graph $\mathcal{G}$, in which the directed edge $(i,j)\in V\times V$ belongs to $E$ if and only if $i\in\pa(j)$. 
\end{definition}
An extreme node observation $X_j$ in the DAG $\mathcal{G}$ is
either the result of an extremely (external) noise $\varepsilon_j$, or the result of (weighted) sum
of observations from the parents of $j$ in $\mathcal{G}$.  
%In a LSCM over $X_1,\dots,X_d$ and associated DAG $G=(V,E)$, we say that $X_i$ causes $X_j$, if $i\in\an(j,G)$. 
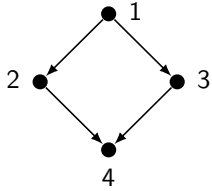
\begin{figure}
    \centering
    \[
\begin{tikzcd}[mdiagram]
 & |[dot,label=right:\textsf{1}]| \arrow[dl] \arrow[dr]& \\
 |[dot,label=left:\textsf{2}]| \arrow[dr] & & 
    |[dot,label=right:\textsf{3}]| \arrow[dl]\\
 & |[dot,label=below:\textsf{4}]| & \\
\end{tikzcd}\qquad
\]
    \caption{Diamond-shaped DAG leading example as in \cite{Gissibl_2018}, with $\mathcal{G}=(V,E)$ where $V=\{1,2,3,4\}$ and $E=\{(1,2),(1,3),(2,4),(3,4)\}$. }
    \label{ch19:fig:diamond}
\end{figure}

\begin{example}[Diamond Example]
\label{ch19:expl:diamond}
We consider the DAG of Figure~\ref{ch19:fig:diamond} where each node $i\in V$ represents a random variable $X_i$. Suppose that the  joint distribution of $\mathbf{X} = (X_1,X_2,X_3,X_4)^\top$ is induced by the LSCM system \eqref{ch19:SCM} with equations:
\begin{equation*}
X_1 = \varepsilon_1; X_2 = \beta_{21}X_1 + \varepsilon_2; X_3 = \beta_{31}X_1 + \varepsilon_3; X_4 = \beta_{42}X_2 + \beta_{43}X_3 +\varepsilon_4,
\end{equation*}
where the coefficients $\beta_{21}, \beta_{31}, \beta_{42}, \beta_{43} \in\Reals\setminus\{0\} $ and $ \varepsilon_j$, $j=1,\ldots,4$ are jointly independent noise variables. An alternative representation of $\mathbf{X}  = (X_1,X_2,X_3,X_4)^\top$ is 
\begin{equation*}
X_1 = \varepsilon_1; X_2 = \beta_{21}\varepsilon_1 + \varepsilon_2; X_3 = \beta_{31}\varepsilon_1 + \varepsilon_3; X_4 = (\beta_{21}\beta_{42}+ \beta_{31}\beta_{43})\varepsilon_1 +\beta_{42}\varepsilon_2+\beta_{43}\varepsilon_3+\varepsilon_4,
\end{equation*}
which can be re-written in terms of the noises as
\begin{equation*}
X_1 = \varepsilon_1; X_2 = \beta_{21}^{\prime}\varepsilon_1 + \varepsilon_2; X_3 = \beta_{31}^{\prime}\varepsilon_1 + \varepsilon_3; 
X_4 = \beta_{41}^{\prime}\varepsilon_1 +\beta_{42}^{\prime}\varepsilon_2+\beta_{43}^{\prime}\varepsilon_3+\varepsilon_4.
\end{equation*}

\end{example}

In general, any LSCM can be written under the linear form
\begin{equation}
\label{ch19:LSCMnoise}
    X_j=\sum_{i\in \An(j)} \beta_{ji}^{\prime}\varepsilon_i,
    %\sum_{i\in \an(j)} %\beta_{ji}^{\prime}\varepsilon_i+\varepsilon_j, {\rm{or}} 
\end{equation} 
where $\beta_{ji}^{\prime} \in \Reals\setminus\{0\}.$
Thus, every component of a LSCM has a linear representation in terms of its ancestral noise variables plus an independent noise as $\beta^{\prime}_{jj}=1$. The coefficients $\beta_{ji}^{\prime}$ are determined by a path analysis of $\mathcal{G}$. Consider the set of all paths from $i$ to $j$, denoted $P_{ij}$ and an element $p$ of this set, i.e., a path of the form $\left[i = k_0 \rightarrow k_1 \rightarrow \cdots \rightarrow k_n = j\right]$. We define the weight of $p$ as the product of the edge weights $\beta_{k_{l+1}k_l}$, $0\leq l \leq n-1$ along $p$
\begin{equation}  
\label{ch19:eq:dPath}
d_{ij}(p)=\prod_{l=0}^{n-1}\beta_{k_{l+1}k_{l}}.
\end{equation}
We illustrate through our leading example, how the LSCM coefficients $\beta_{ji}^{\prime}$ can be obtained from the paths from node $i$ to node $j$ in $\mathcal{G}$.

\begin{example}[Diamond Example]
Consider the path from node $i=1$ to node $j=4$ of the diamond-shaped DAG of Figure \ref{ch19:fig:diamond}. Then $P_{14}$ is composed of two paths, each of length $n=2$, one being $p_1=\left[k_0 = 1 \rightarrow k_1=2 \rightarrow k_2 = 4\right]$ and the other one $p_2=\left[k_0 = 1 \rightarrow k_1=3 \rightarrow k_2 = 4\right]$, with respective weights $d_{14}(p_1)=\beta_{11}\beta_{21}\beta_{42}$ and  
$d_{14}(p_2)=\beta_{11}\beta_{31}\beta_{43}.$
Under our setting, we have that $\beta_{jj}=1$ for all $j\in V$, and we note that 
$$d_{14}(p_1)= \beta_{21}\beta_{42}; d_{14}(p_2)= \beta_{31}\beta_{43},$$
and $\beta^{\prime}_{41} = d_{14}(p_1)+ d_{14}(p_2).$
\end{example}

In terms of noises \eqref{ch19:LSCMnoise}, the coefficients $\beta^{\prime}_{ij}$ are 
$$\beta^{\prime}_{ji} = \sum_{p \in P_{ij}}d_{ij}(p),$$ for $i \in \an(j)$, $\beta^{\prime}_{ji} = 0$, for $i \in V$ \textbackslash $\An(j)$. 
Without loss of generality and for the remainder of this paper, we consider that the DAG is well-ordered, that is that $V$ is topologically ordered in a way compatible with $\mathcal{G}$ such that $k \in \pa(j)$ implies that $k < j$. An implication of this order is illustrated below. 
\begin{example}[Diamond Example]
The DAG in Figure \ref{ch19:fig:diamond} is well-ordered and the coefficients of the LSCM form in terms of noises \eqref{ch19:LSCMnoise} can be represented by the matrix 
\begin{equation}
\label{ch19:MatrixBLSCM}
%\mathcal{\beta}^{\prime}=
\left[\begin{array}{cccc}
   1   &\beta_{21}&\beta_{31}&\beta_{21}\beta_{42}+ \beta_{31}\beta_{43}\\
    0 &1 &0&\beta_{4,2}\\
    0 & 0 & 1 & \beta_{43} \\
    0 & 0 & 0 & 1
\end{array}\right].
\end{equation}
where the $ij$-entry corresponds to $\beta^{\prime}_{ji}$ of \eqref{ch19:LSCMnoise}. In a similar way, a matrix will be determined for the max-linear models defined in \ref{ch19:df:maxlin}, with the sum replaced by the maximum leading to important results for max-linear models.  \end{example}
LSCM for extremes have been studied in \cite{Gnecco_2021}. 
A LSCM for extremes supposes that the noise variables $\varepsilon_1,\dots,\varepsilon_d$ are heavy-tailed, and more precisely that they are regularly varying with index $\alpha >0$, denoted $\varepsilon_1,\dots,\varepsilon_d \in \RV_{\alpha}$. This implies that the random variables $X_1,\dots,X_d$ are independent regularly varying with comparable upper tails. In other words, that there exist $c_1,\dots,c_d>0$ and $\ell$ a slowly varying function such that, for each $j\in\{1,\dots,d\}$, $\Prob(X_j>x)\sim c_j\ell(x)x^{-\alpha}$ as $x\to\infty$. For more details about regularly varying causal linear processes, see \cite{Mikosch_book}, chapters 8 and 11 and \cite{davis_85} for (causal) moving average processes with the form \eqref{ch19:LSCMnoise}, where the causality results from the linear ordering of the underlying DAG. The assumption of regularly varying noise variables with same index $\alpha$ allows to distinguish between the different possible causal relationships between two random variables $X_i$ and $X_j$. The linearity of the LSCM \eqref{ch19:SCM} in the context of extremes finds its justification through the fact that in practice, causal mechanisms may simplify in the tails. Under this linearity, the non-Gaussianity actually helps to identify the causal structure (see, Section \ref{ch19:sct:identifiabiliy}), similarly to the LINGAM method \cite{lingam} in the context of average values. 

While LSCM are based on sum, recursive max-linear models introduced by \cite{Gissibl_2018} use maxima for which the distributions of the noise variables are extreme value distributions or distributions in their maximum domain of attraction. The idea of such models is to let the extremes propagate throughout a network. 
\begin{definition}\label{ch19:df:maxlin}
Consider a set of random variables $\mathbf{X}$. The following relation
\begin{equation}
\label{ch19:MaxL}
X_j := \max_{k \in \pa(j)} \max( c_{kj} X_k, c_{jj} \varepsilon_j),
\quad  j\in V, 
\end{equation}
  with independent non-negative random variables $\varepsilon_1,\ldots,\varepsilon_d$ and strictly positive weights $c_{kj}$ for all $j\in V$ and $k\in \pa(j)\cup \{j\}$, defines a recursive max-linear model (RMLM) with associated directed acyclic graph $\mathcal{G}$.  The vector $\boldsymbol{\varepsilon}= (\varepsilon_1,\ldots,\varepsilon_d) \in \mathbb{R}^d_{+}$ is the vector of innovations.
\end{definition}
Therefore, an extreme node observation $X_j$ in the DAG $\mathcal{G}$ is
either the result of an extremely (external) innovation $\varepsilon_j$, or the result of the maximum
of (weighted) observations from the parents of $j$ in $G$. Without loss of generality, we set $c_{jj}=1$ throughout the paper, to remain consistent with the notations introduced for the LSCM.

\begin{example}[Diamond Example]
As for the LSCM, there is a max-linear model \eqref{ch19:MaxL} associated to the DAG of Figure \ref{ch19:fig:diamond}. Suppose that the  joint distribution of $\mathbf{X} = (X_1,X_2,X_3,X_4)^\top$ is induced by the RMLM system \eqref{ch19:df:maxlin} with equations:
\begin{equation*}
X_1 = \varepsilon_1; X_2 = \max(c_{12}X_1, \varepsilon_2); X_3 = \max(c_{13}X_1 ,\varepsilon_3); 
X_4 = \max(c_{24}X_2,c_{34}X_3,\varepsilon_4).
\end{equation*}
Similarly to the LSCM case, the equations can be reformulated as a max-linear model in terms of noises, leading to the equations system
\begin{equation*}
X_1 = \varepsilon_1; X_2 = \max(b_{12}\varepsilon_1, \varepsilon_2); X_3 = \max(b_{13}\varepsilon_1,\varepsilon_3); 
X_4 = \max(b_{14}\varepsilon_1, b_{24}\varepsilon_2,b_{34}\varepsilon_3,\varepsilon_4),
\end{equation*}
where $b_{12}=c_{12}$; $b_{13}=c_{13}$;  $b_{14}=\max(c_{12}c_{24},c_{13}c_{34})$; $b_{24}=c_{24}$, and $b_{34}=c_{34}$.
As for the LSCM case, the max-linear coefficients $b_{kj}$ are the components of the matrix $B$ which, in this case, is
\begin{equation}
\label{ch19:MatrixB}
B=\left[\begin{array}{cccc}
   1   &c_{12}&c_{13}&\max(c_{12}c_{24},c_{13}c_{34}) \\
    0 &1 &0&c_{24}\\
    0 & 0 & 1 & c_{34} \\
    0 & 0 & 0 & 1
\end{array}\right].
\end{equation}
\end{example}
This leads to the important result due to \cite{Gissibl_2018} that the general recursive max-linear model \eqref{ch19:MaxL}, also called max-linear Bayesian network, can be written 
as 
\begin{equation}
\label{ch19:MaxLM}
X_j := \max_{k =1}^d b_{kj} \varepsilon_k,
\quad  j\in V, 
\end{equation}
with innovations $\varepsilon_1,\ldots,\varepsilon_d$ as in \eqref{ch19:MaxL} and $B=\left(b_{ij}\right)_{d\times d}$ a matrix (the Kleene star matrix in tropical algebra) with non-negative entries. Precisely, the entries of the matrix $B$ also termed the max-linear (ML) coefficients are
\begin{equation}
    \label{ch19:eq:bij}
    b_{ij} = \max_{p \in P_{ij}}d_{ij}(p),
\end{equation} for $i \in \an(j)$, $b_{ij} = 0$, for $i \in V$ \textbackslash $\An(j)$. Here, $d_{ij}(p)$ refers to the path weight defined in \eqref{ch19:eq:dPath} and is re-written in terms of the coefficients $c_{kj}$ of \eqref{ch19:MaxL} as 
$$d_{ij}(p)= \prod_{l=0}^{n-1}c_{k_lk_{l+1}}.$$ 
The ML coefficients are important for identifiability of the RMLM as we will see in Section \ref{ch19:sct:identifiabiliy}.

The distribution of the random vector $\mathbf{X}$ is characterized by the distribution of the innovations $\boldsymbol{\varepsilon}$ and the ML matrix $B$. When the innovations are regularly varying with index $\alpha \in \Reals_{+}$, there exists a normalizing sequence $a_n \in \Reals_{+}$ such that for $n$ independent copies $\mathbf{X}^{(1)},\ldots,\mathbf{X}^{(n)}$ of $\mathbf{X}$ $$\frac{1}{a_n}\max_{s=1}^n \mathbf{X}^{(s)} \rightarrow^d \mathbf{M},\quad n\rightarrow \infty,$$
where the max operator applies componentwise and the limiting vector $\mathbf{M}$ is max-stable. Thus, under the RMLM setting, regular variation of the innovations is transferred to the nodes, i.e., if $\boldsymbol{\varepsilon} \in RV_{\alpha}$, then $\mathbf{X}$ defined by \eqref{ch19:MaxL} is also regularly varying with index $\alpha$. It can be shown that the limiting vector $\mathbf{M}$ is again a RMLM on graph $\mathcal{G}$, with same weights than those of $\mathbf{X}$ in \eqref{ch19:MaxL} with standard $\alpha$-Fr\'echet distributed innovation variables. 

As any multivariate max-stable distribution can be approximated arbitrarily well by a max-linear model (see, e.g., \cite{wang_stoev_2011}), such models became a useful object in the study of causality at extreme levels.

% regularly varying and follows a non-degenerate distribution $H$ with discrete spectral measure.  Explicit formulas for $H$ and its univariate and bivariate margins can be found in Proposition A.2 of \cite{Gissibl_2018}. 

%\textcolor{blue}{TODO: write here on causal ordering + link between the tail dependence matrix and standardized max-linear coefficient
%matrix. See Section 3.4 of the Master thesis of Krali} \\
%\textcolor{red}{Not sure we should mention that link here as it is directly linked to what Mario and Claudia propose for structure learning.}

\subsection{Identifiability of the models}
\label{ch19:sct:identifiabiliy}
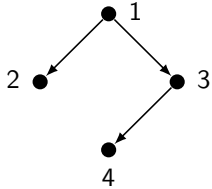
\begin{figure}
    \centering
    \[
\begin{tikzcd}[mdiagram]
 & |[dot,label=right:\textsf{1}]| \arrow[dl] \arrow[dr]& \\
 |[dot,label=left:\textsf{2}]|  & & 
    |[dot,label=right:\textsf{3}]| \arrow[dl]\\
 & |[dot,label=below:\textsf{4}]| & \\
\end{tikzcd}\qquad
\]
    \caption{Diamond-shaped DAG modified example with $\mathcal{G}^B=(V,E^B)$ where $V=\{1,2,3,4\}$ and $E^B=\{(1,2),(1,3),(3,4)\}$. }
    \label{ch19:fig:diamondcut}
\end{figure}

In this section, by ``identifiability", we mean addressing the question of identifiability of the coefficients of the SCM and the associated DAG from the observational distribution $\mathcal{L}(\mathbf{X})$ of $\mathbf{X}$.
%While a LSCM may capture accurately certain causal relationships of extremes, it is not always faithful. Indeed, if the true causal relationships in the system are nonlinear, a LSCM fails to faithfully represent the data.
%Also, if the estimation procedure does not estimate these relationships in an accurate way, the resulting model  may not faithfully represent the true causal structure of the system. \textcolor{red}{not sure this last sentence makes sense here. }
In SCM, we rely either on nonlinearity of the functional form and get identifiability even in the Gaussian case (noise), or on the non-Gaussianity of the noise to get identifiability \cite{lingam}. This includes the case of identifiability of the LCSM for extremes.
%In LSCM with Gaussian noise, the graph can be identified from the joint distribution only up to Markov equivalence classes, assuming faithfulness. 
In the Gaussian case, full identifiability is obtained when all noise variables have the same variance \cite{Peters_2014}.
%Identifiability of Gaussian structural equation models with equal error variances, Biometrika (2014) 
\\
Identifiability of RMLM is studied in \cite{Gissibl_2021} and \cite{Kluppelberg_Lauritzen}. Consider a path from node $i$ to node $j$ and the ML coefficients which are the entries of the matrix $B$, where, for distinct $i,j \in V$, $b_{ij}$ \eqref{ch19:eq:bij} is positive if and
only if there is a path from $i$ to $j$. This
information is contained in the reachability matrix $R = (r_{ij})_{d\times d}$ associated to $\mathcal{G}$, with entries
$r_{ij} = 1$ if there is a path from $i$ to $j$, or if $i = j$,
and 0, otherwise.
If the $ij^{\rm{th}}$ entry of $R$ is equal to one, then $j$ is reachable from $i$. A path from $i$ to $j$ is a max-weighted path if its weight is the maximum, that is $b_{ij}$. As we will see below, the max-weighted paths concept is important in identifiability. 
In many situations, there may exist several DAGs and sets of weights $c_{kj}$ that lead to a random vector $\mathbf{X}$ satisfying \eqref{ch19:MaxL}. It is therefore generally not possible to identify the true DAG and the set of
weights $c_{kj}$ underlying $\mathbf{X}$ of \eqref{ch19:MaxL} from the distribution $\mathcal{L}(\mathbf{X})$. The smallest DAG denoted $\mathcal{G}^B$ and called {\it the minimum ML DAG} of $\mathbf{X}$ is the DAG that has an edge from $i$ to $j$ if and only if this is the 
only max-weighted path from $i$ to $j$. 
\begin{example}[Diamond Example]
Assume that $c_{12}=c_{13}=c >0$, for simplicity of illustration. Our RMLM equations are 
\begin{equation*}
X_1 = \varepsilon_1; X_2 = \max(cX_1, \varepsilon_2); X_3 = \max(cX_1 ,\varepsilon_3); 
X_4 = \max(c_{24}X_2,c_{34}X_3,\varepsilon_4).
\end{equation*}

If $c_{24} \leq c_{34}$, then any DAG $\mathcal{G}^\star$ that is a copy of $\mathcal{G}$ with weight $c_{24}^{\star}$ such that $c_{24}^{\star} \in [0,c_{34}]$, verifies $b_{14}=c_{13}c_{34}$ and the associated representation $X_4 =  \max(c_{24}^{\star}X_2,c_{34}X_3,\varepsilon_4)$
does not alter the distribution $\mathcal{L}(\mathbf{X})$ of $\mathbf{X}$. In this case, $\mathbf{X}$ follows the recursive ML model on the DAG $\mathcal{G}^{\star}$ with set of edge weights $c_{12}=c_{13},c_{24}^{\star}$, and $c_{34}$. But $\mathbf{X}$ also follows a recursive ML on the minimum ML DAG $\mathcal{G}^{B}$ which has the form provided in Figure \ref{ch19:fig:diamondcut} and with edge weights $c_{12}=c_{13}$ and $c_{34}$. This means that one cannot identify $\mathcal{G}$ and its edge weights from $\mathcal{L}(\mathbf{X})$ of $\mathbf{X}$. 
\end{example}

Note that the ML coefficients in $B$ are uniquely determined and 
the following theorem summarizes important results based on $\mathcal{G}^B$.
\begin{theorem}
\label{ch19:thm1}
(Theorem~2 of \cite{Gissibl_2021}). 
Suppose $\mathbf{X}$ follows a recursive ML model \eqref{ch19:MaxL} with edge weights summarized in the matrix $C = c_{ij}$ and ML coefficient
matrix $B$. Let $\mathcal{G}^B$ be the minimum ML DAG of $\mathbf{X}$ as described above. Then a DAG $\mathcal{G}^{\star}$ with associated
edge weights matrix $C^{\star}$ is a valid representation of $\mathbf{X}$ if and only if
\begin{itemize}
    \item[(a)] $\mathcal{G}^B \subseteq \mathcal{G}^{\star}$;
    \item[(b)] $\mathcal{G}^B$ and $\mathcal{G}^{\star}$ have the same reachability matrix $R$;
    \item[(c)] $c^{\star}_{kj} = c_{kj}$ for $k \in \pa^B(j)$;  
    \item[(d)] $c^{\star}_{kj} \in (0,b_{kj}]$ for $k\in \pa^{\star}(j)$ \textbackslash $\pa^B(j)$

\end{itemize}
where $\pa^{B}(j)$ and $\pa^{\star}(j)$  denote the parents of $j$ in $\mathcal{G}^B$ and $\mathcal{G}^{\star}$, respectively.
\end{theorem}
In other words, the only uniquely determined edge weights $c_{kj}$ in the representation \eqref{ch19:MaxL} of $\mathbf{X}$ are those contained in the DAG $\mathcal{G}^B$, that is by $b_{kj}$ of \eqref{ch19:MaxLM} otherwise, $c_{kj}$ can be any strictly positive value smaller than or equal to $b_{kj}$. 
The identifiability of the whole class of DAGs
and edge weights representing the RMLM \eqref{ch19:MaxL} of $\mathbf{X}$ from $\mathcal{L}(\mathbf{X})$ is based on the results exposed in Theorem~(\ref{ch19:thm1}). Clarifying that this class of DAGs can
be recovered from $B$ and showing that $B$ is identifiable from $\mathcal{L}(\mathbf{X})$, \cite{Gissibl_2021} stated the main result on identifiability of RMLM through the following theorem.

\begin{theorem}
(Theorem~1 of \cite{Gissibl_2021}). Let $\mathcal{L}(\mathbf{X})$ be the distribution of $\mathbf{X}$ following a recursive ML model. Then, its ML
coefficient matrix $B$ and the distribution of its innovation vector $\boldsymbol{\varepsilon}$ are identifiable from $\mathcal{L}(\mathbf{X})$.
Furthermore, the class of all DAGs and edge weights that could have generated $\mathbf{X}$ by (\ref{ch19:MaxL}) can be
obtained.
\end{theorem} 
Finally, due to the identifiability of $B$ from $\mathcal{L}(\mathbf{X})$ and the independence of the innovation vector $\boldsymbol{\varepsilon}$, the distribution of the innovation vector $\boldsymbol{\varepsilon}$ is also identifiable from $\mathcal{L}(\mathbf{X})$.
%More about structural properties as well as graph properties %of the recursive ML model for $\mathbf{X}$ can be found in %\cite{Gissibl_2018,Gissibl_2021}.

A more realistic statistical model has been introduced by \cite{Buck_2021} incorporating to the RMLM some random observational noise. More precisely, the noisy model extends \eqref{ch19:MaxL} to a RMLM with propagating noise defined as 
\begin{equation}
\label{ch19:MaxL_Noise}
U_j := \big\lbrace\max_{k \in \pa(j)} \max( c_{kj} X_k, c_{jj} \varepsilon_j)\big\rbrace Z_j,
\quad  j \in V,
\end{equation}
where the i.i.d. noises $Z_j \geq 1$ are atom-free random variables that are unbounded above and independent of the innovation vector $\boldsymbol{\varepsilon}$. 
As for the non-noisy model \eqref{ch19:MaxL}, the question of identifiability of the DAG $\mathcal{G}$ and the edge weights $c_{kj}$ of the ML model with recursive noise from the distribution of $\mathbf{U}=(U_1,\ldots,U_d)^\top$ arises. Similarly to the non-noisy model \eqref{ch19:MaxL}, the true DAG $\mathcal{G}$ and the edge weights underlying $\mathbf{U}$ in representation \eqref{ch19:MaxL_Noise} are generally not identifiable from the distribution of $\mathbf{U}$.  However, \cite{Buck_2021} showed that, again similarly to the non-noisy case, the ML coefficient matrix $B$ is identifiable
from the distribution of $\mathbf{U}$. A consequence of this is that the
minimum ML DAG $\mathcal{G}^B$ is also identifiable and one can also identify the
class of all DAGs and edge weights that could have generated $\mathbf{U}$ (see, \cite{Buck_2021}, Section~4). However, one cannot identify innovations $\boldsymbol{\varepsilon}$ 
or noise variables $\mathbf{Z}$. 

%Although max-weighted paths allow to encode conditional independence relations in the distribution of $\mathbf{X}$ (see \cite{Gissibl_2018}), recursive max-linear models are rarely faithful \cite{Kluppelberg_Lauritzen}, meaning that the global Markov property do not encode all conditional independence among $X_1,\dots,X_d$. In other words, the usual d-separation criterion on the DAG may not identify all valid conditional independence relations. In this sense, max-linear (Bayesian) networks are different from standard Bayesian networks based on linear structural equations or on discrete random variables. \cite{Amendola_2022} studied the Markov properties of a max-linear Bayesian network based on a new separation concept describing conditional independence relations. 
\section{Causal structure learning for extremes}
\label{ch19:sct:Learning}
We now turn our attention to the second task of causal inference, namely, structure learning or causal discovery. 

\subsection{Score-based causal discovery}
Here, we review a class of causal discovery methods that exploits the asymmetry between the cause and effect. Causal asymmetry is the result of the principle that an event is a cause only if its absence would not have been a cause. From there, uncovering the causal direction becomes a matter of comparing a well-defined score in both directions \cite{Mooij_2016}. There are two venues for constructing the causal score. First, the causal score can be model-agnostic. Reflecting the stability principle of \cite{pearl_09}, one can define a measure of independence between the effect and the cause conditional on the effect. \cite{Mhalla_2020} pursued this venue by measuring independence through Kolmogorov complexities. Their approach CausEv relies on an algorithmic formalisation of the notion of stability that yields an asymmetry in the amount of shared information between the cause (say $X$) and the effect given the cause (say $Y \mid X$), and the effect $Y$ and the cause given the effect $X \mid Y$. Through the minimum description length principle of \cite{Rissanen1989}, they show that this asymmetry induced by the presence of causal effects between two random variables $X$ and $Y$ can be translated in terms of inequality in quantile scores where suitable distribution functions of $X$ and $Y$ are defined. Since their method is targeted at discovering causality in extremes, they apply the minimum description length principle with respect to the class of extreme value distributions where a GP distribution is assumed for the margins in the upper quadrant region of the data and an extreme value copula describes their dependence. Thus, the assessment of the direction of causality (in a bivariate setting) becomes a matter of comparing quantile scores in extreme regions, namely the upper quadrant region. Denoting by $\hat{S}_{X^{\text{ext}}}(\tau)$ the $\tau$-th quantile score of the extreme counterpart of a random variable $X$, i.e., $X \mid X> u$ for a high threshold $u$, \cite{Mhalla_2020} define the causal score as
\begin{equation}
S^{\text{ext}}_{X \rightarrow Y} = \dfrac{\hat{S}_{Y^{\text{ext}}}(\tau) + \hat{S}_{X^{\text{ext}} \mid Y^{\text{ext}}}(\tau)}{\hat{S}_{X^{\text{ext}}}(\tau)+\hat{S}_{Y^{\text{ext}} \mid X^{\text{ext}}}(\tau)+\hat{S}_{Y^{\text{ext}}}(\tau)+\hat{S}_{X^{\text{ext}} \mid Y^{\text{ext}}}(\tau)}
\label{ch19:eq:CausEv}
\end{equation}
and conclude that $X$ causes $Y$ whenever the score is (strictly) greater than $0.5$. The choice of the quantile $\tau \in (0,1)$ is arbitrarily and the authors propose an integrated quantile score as the causal direction is expected to be stable with respect to the severity of the extremes. The CausEv method makes no assumption on the causal structure. Its validity relies solely on the validity of the asymptotic models for extremes, i.e., the GP distribution modeling the univariate tail behaviour and the extreme value copula describing the tail dependence.

The second venue for constructing the causal score relies on a model-induced asymmetry. This venue was followed by \cite{Gnecco_2021} with the LSCM (see Definition~\ref{ch19:df:scm}) with heavy-tailed noise variables, and \cite{Tran_2022} with the RMLM (see Definition~\ref{ch19:df:maxlin}) supported on a tree structure. Both their approaches are detailed below.

We start with an LSCM-induced graph $\mathcal{G}$, as described by the linear form \eqref{ch19:LSCMnoise}. We denote by $F_k$ the cumulative distribution function of node $X_k$ and assume the noise variables to be independent and regularly varying with index $\alpha>0$. \cite{Gnecco_2021} propose to capture the causally-induced asymmetry in the graph $\mathcal{G}$ by looking at the signal in the bivariate tails through the so-called \textit{causal tail coefficient}
\begin{eqnarray}
    \Gamma_{jk} &=& \underset{u \to 1^{-}}{\lim} \E \{ F_k(X_k) \mid F_j(X_j) > u \} \label{ch19:eq:gamma} \\
    &=& \dfrac{1}{2} + \dfrac{1}{2} \dfrac{\sum_{h \in An(j) \cap An(k)} \beta'_{hj}}{\sum_{h \in An(j)}\beta'_{hj}} \notag, 
\end{eqnarray}
defined for $j \neq k$. Then, a comparison of the coefficients $\Gamma_{ij}$ and $\Gamma_{ji}$ allows to establish the presence (or not) of a causal relationship between nodes $X_j$ and $X_k$ as well as its direction. Precisely, when $X_i$ causes $X_j$, we expect $\Gamma_{ij}$ to be equal to one and $\Gamma_{ji} \in (1/2,1)$. 

\begin{example}[Diamond Example]
    
For the DAG described in Figure~\ref{ch19:fig:diamond}, we compute the causal tail coefficients for nodes $X_1$ and $X_4$. Since $An(1)=\{1\}$ and $An(4)=\{1,2,3,4\}$, we obtain that $\Gamma_{14}=1$ whereas 
$$\Gamma_{41}= \dfrac{1}{2} + \dfrac{1}{2} \dfrac{\beta'_{14}}{\beta'_{44} + \beta'_{34} + \beta'_{24} + \beta'_{14}} < 1.$$
The same goes for nodes $X_1$ and $X_3$ where 
\begin{eqnarray}
    \Gamma_{13} &=& \dfrac{1}{2} + \dfrac{1}{2} \dfrac{\beta'_{11}}{\beta'_{11}} = 1, \notag \\
    \Gamma_{31} &=& \dfrac{1}{2} + \dfrac{1}{2} \dfrac{\beta'_{13}}{\beta'_{13}+\beta'_{31}} < 1. \notag
\end{eqnarray}
\end{example}

At the level of the graph $\mathcal{G}$, the task of structure learning is synonym of retrieving the topological (causal) order and \cite{Gnecco_2021} propose an algorithm called EASE that returns the causal order of a graph based on the matrix of its causal tail coefficients. At the sample level, statistical inference relies on a nonparametric estimator of $\Gamma_{ij}$ which is consistent under mild conditions on the tail behaviour of the marginals $F_j$ and $F_k$ as well as the effective sample size in the tails. Consistency of the algorithm that is achieved when the probability of making a mistake vanishes, follows from the consistency of the estimator of $\Gamma_{ij}$. The methodology developed by \cite{Gnecco_2021} requires the noise variables in the LSCM to have the same tail coefficient, though they empirically show that the method is still valid if the ancestor has lighter tails than its descendants. Additionally, although the authors show that their algorithm is asymptotically robust to hidden confounders, the assumption of comparable tails (of the hidden confounders and the observed variables) is still needed. \cite{Pasche_2022} extended the methodology by conditioning on the values of observed confounders when computing the causal tail coefficient of \cite{Gnecco_2021}. Using a semi-parametric estimator of the marginal distribution functions, \cite{Pasche_2022} propose a parametric \textit{GPD causal tail coefficient} estimator that is able to reduce or remove the effect of potential confounders on the measure of asymmetry $\Gamma_{ij}$.

Following a similar train of thought, \cite{Tran_2022} look at a different measure of asymmetry under the RMLM \eqref{ch19:MaxLM} supported on a root-directed tree $\mathcal{T}$. The proposed measure quantifies the concentration of the distribution of the ratio of two nodes, conditional on one node being extreme. Precisely, working on the logarithmic scale and conditioning on observations exceeding their $\alpha$th quantile ($\alpha$ being a large quantile level) in one margin, the causal score is defined as
$$w_{ij} (r) := \dfrac{1}{n_{ij}} [ \E\{\mathcal{X}_{ij}(\alpha) \} - Q_{\mathcal{X}_{ij}(\alpha)}(r) ]^2,$$
where $\mathcal{X}_{ij}(\alpha) = \{ X_i - X_j: X_j > Q_{X_j}(\alpha)\}$, $n_{ij}= \vert \mathcal{X}_{ij}(\alpha) \vert$, and $Q_{Y}(\tau)$ is the $\tau$th quantile of random variable $Y$. This causal score serves then as a weight for the Chu–Liu/Edmonds' algorithm that is used to obtain a minimum root-directed spanning tree. Consistency of the methodology is proven under assumptions on the signal to noise ratio of the underlying RMLM. While the proposed algorithm is targeted for causal structures described by a RMLM, it heavily relies on properties of the tree structure of the graph. We now focus on a general approach to causal discovery for RMLMs supported on a DAG.

\subsection{RMLM-based causal discovery}
\label{ch19:sec:RMLMdisco}
Recursive max-linear models are a natural tool to characterize causal relations in a system of variables. For instance, RMLMs reflect the intuitive concept of the large shocks propagating through a network and having a dominant effect on their descendants. Additionally, as the size of the network grows, dependence structures under RMLMs remain tractable. Such property is desirable in extreme value theory that revolves around the notion of multivariate regular variation. 

Under the RMLM setting, the task of learning the causal structure boils down to identifying the ML coefficient matrix $B$, with entries defined in \eqref{ch19:eq:bij}. By construction, entries of the ML coefficient matrix represent the weight of the max-weighted paths between two vertices. Hence, the matrix $B$ is closely related to the tail dependence structure of the network. For instance, \cite{Gissibl_2018b} show that if $\mathbf{X}$ follows a recursive ML model with i.i.d. regularly varying innovations with index $\alpha \in \mathbb{R}$, then $\mathbf{X}$ is in the max-domain of attraction of the max-stable distribution $G_{\mathbf{X}}$ given by
\begin{equation}
    G_{\mathbf{X}}(\mathbf{x}) = \exp \bigg \lbrace - \sum_{j=1}^d \max_{i \in De(j)} (b_{ji}/x_i)^{\alpha} \bigg\rbrace, \quad \mathbf{x}=(x_1, \ldots, x_d)^\top \in \mathbb{R}^d_{+}. \label{ch19:G_X}
\end{equation}

Building on this fundamental relationship, \cite{Gissibl_2018b} show that the bivariate tail dependence coefficient defined as
$$\chi(i,j)= \lim_{u \rightarrow 1} \Pr\{X_j > F^{-1}_{j}(u) \vert X_i > F^{-1}_{i}(u) \},$$
where $F_i^{-1}$ is the generalized inverse of the distribution function of the margin $X_i$, is related to the entries of the matrix $B$ through
$$\chi(i,j)= \sum_{k \in An(i) \cap An(j)} \min(\bar{b}_{kl}, \bar{b}_{kj}).$$
Here, the entry $\bar{b}_{ij}$ is the standardized ML coefficient $b_{ij}$, i.e., 
$$\bar{b}_{ij} := \bigg ( \dfrac{b_{ij}^{\alpha}}{\sum_{k \in An(j)} b_{kj}^{\alpha}} \bigg)^{1/\alpha}.$$

Similarly, \cite{Krali_2021} detail the relationship between the ML coefficients and the so-called scaling parameter defined as
$$\sigma^2_{ij} = \int_{S^d} \omega_i \omega_j dH_{\mathbf{X}}(\boldsymbol{\omega}), \quad \boldsymbol{\omega} =(\omega_1, \ldots, \omega_d)^\top \in S^d,$$
where $H_{\mathbf{X}}$ is the limiting spectral measure of $\mathbf{X}$, defined on the unit simplex $S^d= \{ \boldsymbol{\omega} \in \mathbb{R}^d_{+}: \vert \vert \boldsymbol{\omega} \vert \vert =1 \}$. For instance, under the assumption of regular variation (of index $2$) of the innovations of the RMLM, they derive the limiting spectral measure from \eqref{ch19:G_X} and show that the scalings are given by
$$\sigma^2_{ij} = (B B^\top)_{ij}.$$
The authors then propose a causal structure learning algorithm where the ML coefficient matrix $B$ is constructed recursively by inferring the scalings. As the assumption of regular variation on the innovations yields a discrete spectral measure for the observations, the empirical spectral measure is used to estimate the scalings, resulting in asymptotically normal and consistent estimates of the ML coefficient matrix $B$. 

The structure learning algorithm proposed in \cite{Krali_2021} is valid under the assumption of the underlying DAG being well-ordered, though an additional step of finding the source nodes can be implemented if this does not hold. \cite{Krali_2023} extend this work to include settings with hidden confounders. They suppose that the entire DAG is not observed and discuss conditions under which the causal structure can be retrieved. 

The RMLM discussed so far assumes that extreme observations at the node variables are the result of extreme shocks in the innovations propagating deterministically through the network. Thus, and in contrast with the LSCM defined in \eqref{ch19:SCM}, there are no observation errors in the model, which can be deemed unrealistic. This motivates the work by \cite{Buck_2021} who introduce the max-linear model with propagating noise
    \begin{eqnarray}
        U_i = \bigg( \bigvee_{j \in pa(i)} c_{ij}U_j \bigvee \epsilon_i \bigg) Z_i, \quad i=1,\ldots,d \notag
    \end{eqnarray}
where $Z_i$ are i.i.d. regularly varying noise variables, independent of the innovations $\varepsilon_i$. This model can be shown to have a representation in terms of max-linear random coefficients, i.e., it can be written as \eqref{ch19:MaxLM}. The LM matrix $B$ is random as it also depends on the noise variables. \cite{Buck_2021} use the concept of minimum ratios to estimate the DAG associated to the propagating noise model. The causal structure learning problem is framed as an optimization problem, where the topological order is obtained by minimizing an objective function involving minimum ratios of the nodes, and an estimation problem, where the random ML coefficients are inferred from the estimated order.

One perspective on observational causal discovery for extreme events is in terms of recursive max-linear models. Thanks to the causal interpretation of its parameters as well as its close link to the notion of multivariate regular variation, the structure of such models can be leveraged to learn the causal relations governing a given network. It remains only to decide on the adequacy of the structure.

\subsection{Application to a real dataset}
\label{ch19:sct:Seine}
\begin{figure}[htp]
    \centering
\includegraphics[width=0.9\textwidth]{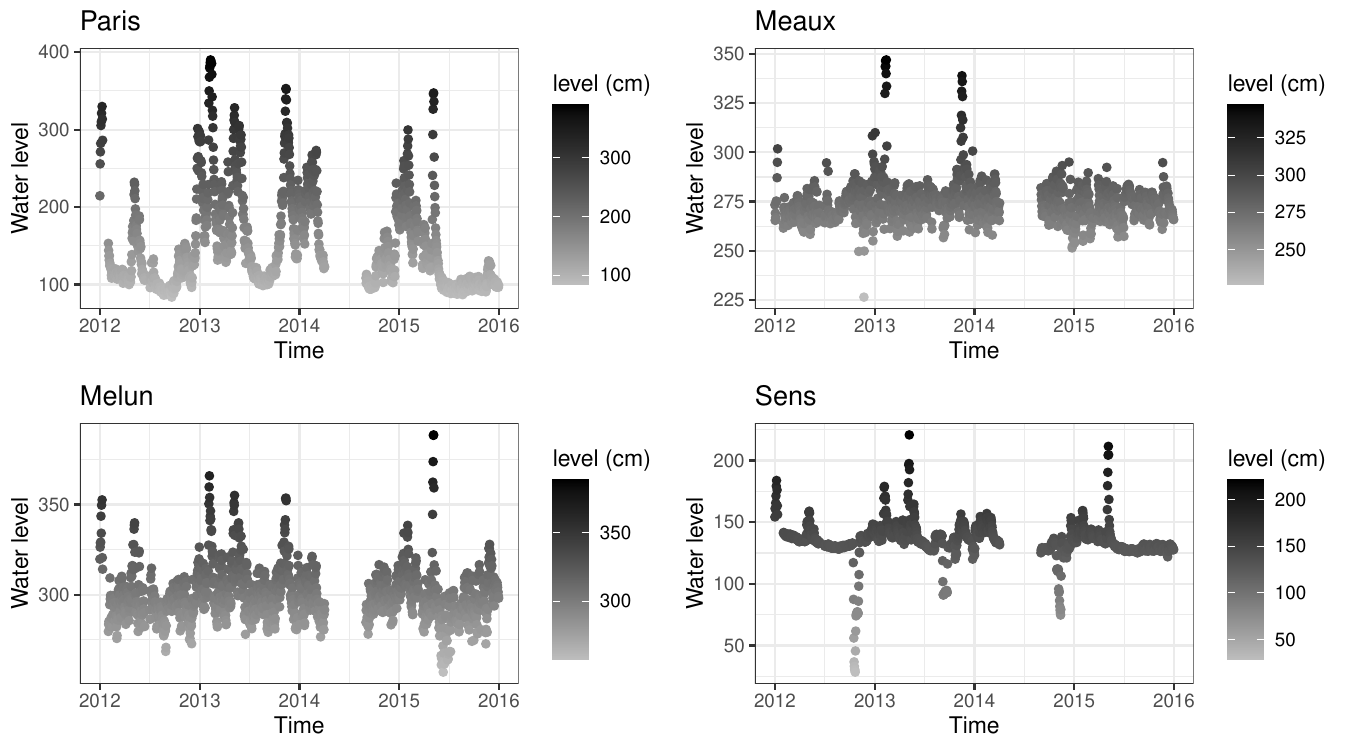}
\caption{Daily water levels at four sites on the Seine network from 2012 to 2015.}
\label{ch19:fig:seine_ts}
\end{figure}

We apply the causal extreme discovery methods described in Section \ref{ch19:sct:Learning} to the Seine network data introduced in Section~\ref{ch19:sec:Seine}. The three methods we consider are the model-agnostic CausEv approach of \cite{Mhalla_2020}, the LSCM-based EASE algorithm of \cite{Gnecco_2021}, and the RMLM-based approach described in \ref{ch19:sec:RMLMdisco}. The data, kindly provided by the authors of \cite{Asenova2021}, come from the website of the French Ministry of Ecology, Energy, and Sustainable Development, specifically \texttt{http://www.hydro.eaufrance.fr}. The dataset consists of $n=27893$ daily water levels in centimeters measured at the five stations and covers the time span from January 1987 to April 2019, with occasional gaps in data for certain measurement stations. Contrary to mountain rivers such as the Danube network \cite{asadi2015,Mhalla_2020,Gnecco_2021} or the Swiss network \cite{Pasche_2022}, where the more extreme observations happen during summer when rivers are less likely to be frozen and flash floods are more frequent, the Seine network does not display any seasonality. For instance, as displayed in Figure~\ref{ch19:fig:seine_ts}, very large observations can occur at any time throughout the year, though these events might be of different types, e.g., fluvial or pluvial. 

\begin{figure}[htp]
    \centering
\includegraphics[width=4cm]{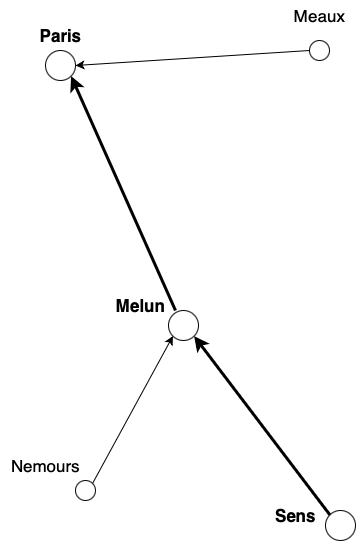}
\caption{Oriented network associated to the Seine river network.}
\label{ch19:fig:SeineDAG}
\end{figure}

The map in Figure \ref{ch19:fig:SeineDAG} shows the oriented network associated to the Seine river in terms of water flow. 

%%%%%%%%%%%%%%%%%%%%%%%%%%%%%%%%%%%%%%%%%%%%%%%%%%%%%%%%%%%%

Since the considered sites are either on the Seine or on one of its tributaries, we expect their characteristics, such as the volume of the associated river, to potentially influence the tail behaviour of the water level variable. Thus, before conducting any causal discovery, we fit a GP distribution to the tail of each series and obtain an estimated value of the shape parameter $\xi$ reported in Table~\ref{ch19:tab:SeineXi}. While the three estimated values for Meaux, Melun, and Sens stations are rather close, the one for Paris is slightly lower compared to the others. All four 95\% confidence intervals for the shape parameter overlap, though. More importantly, the estimated shape parameter of Nemours is very large, violating the tail equality assumption required by the two model-based methods (EASE and the one of Section \ref{ch19:sec:RMLMdisco}). Although this assumption is not required for the CausEv approach, we decided to remove this station when applying the three methods. 

\begin{table}
\begin{tabular}{c|c|c|c|c|c}
\hline
Stations & Paris &  Meaux &   Melun & Nemours  &   Sens  \\
\hline
$\hat{\xi}$ & 0.099(0.11) &0.31(0.13) &0.25(0.12)& 0.87(0.14)& 0.31(0.11) \\
%Declustered & 0.013(0.35) & 0.050(0.27)&  0.0058(0.29) & 0.82(0.41)&  0.081(0.33)\\
\end{tabular}
\caption{Shape parameter estimate ($\hat{\xi}$) obtained from fitting a GP distribution to the tail of the five stations.}
\label{ch19:tab:SeineXi}
\end{table}

Table \ref{ch19:tab:DAGmatrix} shows the reachability matrix associated with the oriented network of Figure~\ref{ch19:fig:SeineDAG} of the Seine without Nemours, where a value of 1 in the cell $(i,j)$ means that there is a path from $i$ to $j$. We apply the three approaches CausEv \cite{Mhalla_2020}, EASE \cite{Gnecco_2021}, and RMLM-based \cite{Krali_2021} on the non-declustered data and the resulting reachability matrices are summarized in Table~\ref{ch19:tab:CausEvmatrix}, Table~\ref{ch19:tab:LSCMmatrix}, and Table~\ref{ch19:tab:RMLMmatrix}, respectively. More specifically, Table \ref{ch19:tab:CausEvScores} shows the 95\% confidence intervals for the CausEv score \eqref{ch19:eq:CausEv} obtained by a bootstrap approach resampling the years. Whenever the value 0.5 is not in the 95\% confidence interval of cell $(i,j)$, there is an edge from $i$ to $j$. Table \ref{ch19:tab:LSCMmatrix} is obtained using the EASE algorithm on the estimated causal tail coefficients \eqref{ch19:eq:gamma}.
%obtained by setting the number of exceedances to $k=23$ which corresponds to $\lfloor n^{0.4}\rfloor$. 
Table \ref{ch19:tab:RMLMmatrix} is obtained by applying the method based on the RMLM and described in Section \ref{ch19:sec:RMLMdisco}. The two reachability matrices obtained from EASE and RMLM-based method are the same. 

% Loing Nemours: https://www.hydro.eaufrance.fr/sitehydro/F4370002/impression-synthese

\begin{table}
\begin{tabular}{c|cccc}
\hline
Actual reachability matrix & Paris &  Meaux &   Melun &    Sens  \\
\hline
Paris &0 &0&0&0 \\
Meaux &1 &0&0&0 \\
Melun &1 &0&0&0 \\
Sens &1 &0&1&0 \\

\end{tabular}
\caption{Actual reachability matrix corresponding to the Seine network without Nemours.}
\label{ch19:tab:DAGmatrix}
\end{table}

\begin{table}
\begin{tabular}{c|cccc}
\hline
CausEv reachability matrix & Paris &  Meaux &   Melun &    Sens  \\
\hline
Paris &0 &0&0&0 \\
Meaux &1 &0&0&0 \\
Melun &1 &0&0&0 \\
Sens &1 &0&0&0 \\
\end{tabular}
\caption{Estimated reachability matrix obtained from the method CausEv described in \cite{Mhalla_2020}.}
\label{ch19:tab:CausEvmatrix}
\end{table}

\begin{table}
\begin{tabular}{c|cccc}
\hline
CausEv scores & Paris &  Meaux &   Melun &    Sens  \\
\hline
Paris & &&& \\
Meaux &$[0.505;0.575]$ &&& \\
Melun & $[0.538; 0.585]$ &$[0.465; 0.521]$&& \\
Sens &$[0.509; 0.567]$ &$[0.401; 0.511]$&$[0.466; 0.540]$& \\
\end{tabular}
\caption{95\% confidence interval for the CausEv scores \eqref{ch19:eq:CausEv} from 300 bootstrap samples obtained by resampling the years.}
\label{ch19:tab:CausEvScores}
\end{table}

\begin{table}
\begin{tabular}{c|cccc}
\hline
EASE reachability matrix & Paris &  Meaux &   Melun &    Sens  \\
\hline
Paris &0 &0&0&0 \\
Meaux &1 &0&1&1 \\
Melun &1 &0&0&0 \\
Sens &1 &0&1&0 \\
\end{tabular}
\caption{Estimated reachability matrix obtained from the EASE algorithm for LSCM \cite{Gnecco_2021}.}
\label{ch19:tab:LSCMmatrix}
\end{table}

\begin{table}
\begin{tabular}{c|cccc}
\hline
RMLM reachability matrix & Paris &  Meaux &   Melun &    Sens  \\
\hline
Paris &0 &0&0&0 \\
Meaux &1 &0&1&1 \\
Melun &1 &0&0&0 \\
Sens &1 &0&1&0 \\
\end{tabular}
\caption{Estimated reachability matrix obtained from the RMLM-based method described in Section \ref{ch19:sec:RMLMdisco} \cite{Krali_2021}.}
\label{ch19:tab:RMLMmatrix}
\end{table}

To quantify the different causal inference statements and different intervention distributions resulting from the three approaches, we use the structural intervention distance (SID) \cite{Peters_2015}. In short, the SID evaluates the distance between the estimated reachability matrix and the actual one. To assess the variability of the SID, we use a bootstrap approach by resampling the years with replacement 300 times. The 95\% percentile CIs for the SID are $[0,4]$ for CausEv, $[0,6]$ for EASE, and $[0,6]$ for the RMLM-based method. The results among the three approaches are rather similar.

While we did not decluster the data prior to the analysis, the question of declustering is still open. Indeed, when applied on the declustered data using the approach described in \cite{asadi2015}, the two model-based approaches (EASE algorithm and RMLM method of Section \ref{ch19:sec:RMLMdisco}) did not perform well whereas the reachability matrix obtained from CausEv remains the same than the one obtained on the non-declustered data. For the two model-based approaches it is not clear whether the larger size of the dataset or its temporal dependence feature (or both) are relevant for the structure learning task.

\section{Conclusion}
\label{ch19:sct:Conclusion}
The field of causality for extremes is an emerging area of statistics, and in this work, we have provided a non-exhaustive review of recent methods for causality of extremes from observational data.
Highly relevant in various domains, such as climate, finance, or epidemiology, where understanding the causes of extreme events is of utmost importance for prediction or risk assessment, some open problems are continuously arising. Some of these challenges are the presence of hidden confounders or hidden nodes. \cite{Krali_2023} have provided necessary and sufficient conditions to disregard hidden nodes in regularly varying RMLM. Another question is the possibility to relax the assumption of similar tail heavyness in model-based approaches and to allow lighter tails of the innovations. Still linked to model-based methods, the bad results in the cases of declustered data compared to non-declustered data, even for large datasets, remains somehow not understandable.

There is a link between graphical models for extremes \cite{engelke20} presented in \cite{Engelke_chapter} and causality. The work of \cite{engelke20} based on densities on undirected graphs can complement the literature on recursive SEMs for extremes (e.g., LSCM and RMLM). For instance, \cite{engelke20} define conditional independence for general (continuous) multivariate extreme value models, which can serve as a tool for causal discovery on DAGs, like in classical Gaussian models. 
%So that, formulating DAGs for RMLM using graphical models for extremes \cite{engelke20} would lead to factorizations for directed graphs. 
This can serve as the foundation for expanding research in causal inference for extremes to encompass continuous extreme value distributions. 

Causal structure learning methods for extremes of temporal data with possible latent variables and non linear relations are still relatively rare, but interesting work in this area includes \cite{bodik23} which introduces a causal tail coefficient for (heavy tailed) time series. The Granger causality test is equivalent to testing whether the coefficients of the past values of a series $X(t)$ in an autoregressive model of another one $Y(t)$ involving past values of both time series are equal to zero. Instead of looking at the Granger causality in mean, one may want to consider a Granger causality test in the realized dynamic extreme quantiles of $Y(t)$. 
A change in the dynamics of the causal relationship can be assessed by assuming piecewise autoregressive segments with the graphical help of extremogram \cite{davis09} and estimating the number of structural breaks that would induce different model coefficients through time. 
These examples represent potential avenues for exploring causality in extreme situations, yet numerous additional possibilities exist such as investigating nonlinear functional relationships and detecting causal covariates, among others.

% \mainmatter

\bibliographystyle{apalike}
\bibliography{main.bib}

\printindex
\cleardoublepage

\end{document}